# Control and calculation of the titanium sublimation pumping speed and re-ionisation in the MAST Neutral Beam Injectors

R McAdams

EURATOM/CCFE Fusion Association, Culham Science Centre, Abingdon, Oxfordshire OX14 3DB, UK

Tel.: +44 (0)1235 464539 E-mail address: roy.mcadams@ccfe.ac.uk

## Abstract

A high pumping speed is required in neutral beam injectors to minimise re-ionisation of the neutral beams. The neutral beam injectors on MAST use titanium sublimation pumps. These pumps do not have a constant pumping speed; their pumping speed depends on the gettering surface history and on both the integrated and applied gas load. In this paper we report a method of maintaining a constant pumping speed, through different evaporation schemes, suitable for operations of the MAST beamline for both short and relatively long beam pulses by measurement of the pressure in the beamline. In addition the MCNP code is then used to calculate the gas profile in the beamline by adjusting the input pumping speed to match the measured pressure. This allows the resulting gas profile to be used for calculation of the re-ionisation levels and an example is given.

## 1. Introduction

In Neutral Beam Injection (NBI) it is vital to keep the level of re-ionisation of the neutral beam to a minimum. Re-ionisation leads to loss of transmission due to deflection of the particles in the stray magnetic fields of the tokamak. These fields can also focus the re-ionised particles causing damage. In extreme cases, gas evolution from the regions where the re-ionised particles strike the walls leads to runaway re-ionisation (beam blocking) and the damage could be catastrophic [1,2,3]. Keeping the re-ionisation levels at acceptable levels requires high pumping speeds and neutral beam injectors use a variety of pumping methods [4].

The MAST neutral beam injectors employ titanium sublimation (or getter) pumps (TSP or TGP) where titanium is evaporated onto a surface. Gas colliding with the surface reacts with the titanium to form low vapour pressure compounds i.e. the gas is pumped. As pointed out by Feist [5], "One disadvantage of using TGP is the poor reproducibility of the data. In the case of hydrogen, pumping speed and capacity are dependent on the base pressure during sublimation and the sublimation rate as well as the whole history of the surface." Thus it is essential to find operational regimes that give acceptably high and constant pumping speeds.

The aim of this paper is to describe how such regimes are achieved in the MAST NBI systems for gas pulses of different duration. In addition the measured pressures are used in conjunction with the MCNP code to calculate the pumping speed and the gas profile in the beamline. The gas profile can then be used to calculate the re-ionisation levels. In Section 2 the MAST NBI systems and in particular the titanium sublimation pumps are described. In Section 3 the method of measurement of the beamline pressure is given along with some example operational experience that shows what operating regimes are required for constant pumping speeds to be obtained. The use of the MCNP code to calculate the gas profiles and an example calculation of the re-ionisation levels is described in Section 4. Finally conclusions are given in Section 5.

## 2. The MAST Neutral Beam Injectors and pumping system

### 2.1 The Neutral Beam Injectors

MAST presently has two identical Neutral Beam Injectors; the South and Southwest beamlines. These injectors are capable of extracting a deuterium ion beam of 65A at 75keV [6,7]. The resulting maximum neutral beam power transmitted into MAST is approximately 2.5MW from each injector. The present beam pulse length is up to 0.5s but an upgrade to MAST will see this extended to 5s [7]. A schematic of the injector is shown in Figure 1. The composite beam leaving the neutraliser enters a vacuum tank containing the magnet used to deflect residual ions from the beam onto the residual ion dumps. The neutral beam then passes through an intertank duct into another vacuum tank containing a calorimeter. With the calorimeter gates closed the profile and power of the beam can be measured. When the calorimeter gates are open the beam can then pass beyond the calorimeter and through a series of three ducts into the MAST vessel.

### 2.2 The Titanium Sublimation Pumps

In each of the magnet and calorimeter tanks are the titanium sublimation pumps (TSP) used to achieve relatively low base pressure of ~ $10^{-7}$ mbar ($10^{-5}$ Pa). There is a small turbo molecular pump on each tank for initial pump down of the system. The design of the TSPs is a direct copy of those used at ASDEX [8]; the same moulds have been used to provide the castings of the plates from which the pumps are made. Figure 2 shows a sketch of the pump. Each pump consists of four chambers, made from three panels. The panels are corrugated to increase the surface area. In each chamber are three filaments (although only one is shown in the diagram) which are the titanium baring wires. By passing a current of ~ 80-90A through each wire for a period of two minutes titanium is evaporated onto the chamber walls. A soft start and end to the evaporation period is used to minimise mechanical stress on the wires. The MAST TSPs are 2/9ths the size of the pumps used in ASDEX where a pumping speed of approximately 700,000 l/s is reported [8] in hydrogen operation. Thus the MAST pumps might be expected to have a pumping speed of ~ 110,000 l/s in deuterium operation. However as mentioned in Section 1, the pumping speed depends on the amount of titanium on the surface, the amount of gas already absorbed and the presence of contaminants such as nitrogen.

## 3. Establishment of operational regimes for the TSPs

### 3.1 Pressure measurement

In order to understand the behaviour of TSPs and to calculate their pumping speed, fast Penning gauges were installed in the South beamline. The existing ion gauges were considered both to be too slow and to not have the ability to measure high enough pressures particularly at low pumping speeds. The gauges used were Pfeiffer 050 gauges which have a response time of ~10ms. One gauge was installed in the magnet tank on the top flange near the TSP. The other gauge was installed on the downstream wall of the calorimeter tank. The positions are discussed in Section

4.2. In order to obtain an absolute pressure measurement for the pumping speed determination an Edwards WRG-S-NW25 calibrated gauge was also installed on the magnet tank.

In order to calibrate the Penning gauges, the beamline was vented slowly to atmospheric pressure with nitrogen from a pressure of ~ $10^{-6}$ mbar and the measurements from the Penning gauges were recorded at intervals along with the absolute pressure reading thus giving a calibration to absolute pressure. The appropriate gauge factors for operation in hydrogen (and deuterium), as supplied by the manufacturer were also applied to the calibration. Figure 3 shows an example of the measured pressure traces for a 1.3s gas only pulse (no extracted beam) from the injector at a gas flow rate of 30 mbar l/s (3 Pa $m^3$/s). This gas throughput is the normal operating point for the injector.

**3.2 Operational regimes to maintain constant pumping speed**

For injection of the neutral beam into the tokamak the pumping speed should be constant and also high enough to cause minimum re-ionisation. In Figure 4, the evolution of the peak pressure as a number of 3.5s gas pulses of 31 mbar l/s throughput is shown for both the magnet tank and calorimeter tank. Data are shown where the titanium evaporation durations are two, three and five minutes. In these data there is no gas entering MAST. Once the initial titanium evaporation has been carried out no further evaporations are made. From Figure 4 it is clear that the peak pressure increases and hence the pumping speed deteriorates for the three data sets as the integrated gas load on the pump increases. Some saturation of the gauge reading in the magnet tank can be seen for the two minute evaporation data. Increasing the evaporation period decreases the peak pressure and slows the rate of pressure increase with gas loading. This reflects the increased capacity of additional titanium on the pump. Whilst such decreases in pumping speed are acceptable for asynchronous operation (no injection of neutral beam into the tokamak) whilst conditioning or checking the injector performance it would not be acceptable for synchronous operation (injection into the tokamak) for the reasons given already. The challenge is then to find an operating regime which allows a constant and a sufficiently high pumping speed.

The method used was to repeatedly fire gas pulses at a throughput of 30 mbar l/s. In between the gas pulses a number of chambers in each pump would undergo titanium evaporation for two minutes. This would allow the identification of a constant pumping speed evaporation regime. MAST previously operated with beam pulse lengths of up to 0.5 seconds. It is presently undergoing an upgrade and in future the NBI system will have beam pulse lengths of up to five seconds and so this procedure was carried out for pulse lengths of 1.3s and 6.5s, the additional period allowing some gas equilibration time for the arc discharge to become established in the ion source.

The data from a series of pulses of 1.3s duration are shown in Figure 5. Similarly to Figure 4, when there is no titanium evaporation between pulses the peak pressure in the two tanks increases. On evaporating a titanium wire in all chambers of each pump, the peak pressure starts to fall with each cycle of evaporation. On switching to evaporation in only one (different) chamber in each pulse the peak pressure stays constant. Changing to evaporation of two chambers in each tank between pulses results in the peak pressure falling in the next pulse. It is clear that evaporation of only one chamber in each tank between pulses is sufficient to maintain a constant peak pressure and hence pumping speed.

The measured peak pressures, for a series of 6.5s duration gas pulses under different evaporation regimes are shown in Figure 6. When all the chambers undergo evaporation the peak pressure falls with each succeeding pulse. Continuing further evaporations of all chambers would appear to have been able to lower the peak pressure further. Switching to evaporations in two chambers allow a constant pressure to be maintain and increasing the evaporations to three chambers between each pulse results in the peak pressure falling again. Notice that for both gas pulse durations the peak pressure in the magnet tank is considerably higher than in the calorimeter tank (as was the case in Figure 5). This is discussed in Section 4.5.

## 4. Calculation of the pumping speed using MCNP

### 4.1 The use of the MCNP code

The calculations of the TSP pumping speeds were carried out using the Monte Carlo N-Particle (MCNP) code [9]. The beamline is represented by a number of cells. Each cell can represent an object or a part of an object (magnet, calorimeter etc) or a region of space in which some parameter such as number density might be calculated e.g. along the beam path. The trajectories of a large number of particles, representing the gas molecules, originating from a source are tracked. The walls of cells representing objects are treated as being reflecting with a diffuse cosine reflection distribution. Unless there are sinks where particles are lost the molecules will bounce around forever. The sinks in the system are the vacuum pumps and the MAST tokamak. Particles entering the cells representing the pumps and MAST are no longer tracked and are considered lost i.e. effectively pumped. In this way all the particles are tracked until they are pumped. MCNP allows the calculation of various "tallies" associated with particles passing through each cell or striking a surface and with careful interpretation, these can be related to a number density in the cell. Tallies are set up for cells defined along the beam direction. These give the number density along the beamline (or in any other space where a cell has been defined). Tallies can also be set up which count the number of particles entering the pumps or the MAST tokamak; this particle accounting ensures that the model does not hold areas where particles can be unintentionally lost.

### 4.2 The MCNP model of the MAST NBIs

The MCNP model of the MAST NBI is shown in Figure 7. The features from the drawing in Figure 1 can be clearly identified. The positions of the fast Penning gauges in the magnet and calorimeter tanks are shown.

The use of MCNP, in tracking individual particles, represents free molecular flow. The conductance of the second stage neutraliser is estimated at ~ 10,000l/s for $D_2$ at 300K. Thus, assuming zero pressure at the exit, the pressure at the neutraliser entrance is ~ $3 \times 10^{-3}$ mbar (0.3 Pa) for a gas throughput of 30 mbar l/s. Using a collision cross-section of $10^{-20}$ m$^2$, the mean free path is estimated at 1.4m which is longer than the neutraliser section in the model. Since the pressure will fall through the neutraliser and downstream, the assumption of free molecular flow seems reasonable. In reality the gas will be hot and the conductance higher in this region. In the model however it is assumed that the gas is at 300K throughout.

The pump in each tank is represented by four rectangular boxes whose height is that of the actual pump chambers. Five of the sides reflect particles but the entrance is transparent. One a particle enters through this face it is no longer tracked and is considered to have been pumped. The pumping speed of the pump in each tank is set by adjusting the width of the strip representing each of the chambers of the pump. The height of the pump is fixed and so the pumping speed, S, is given by

$$S = \frac{1}{4}\bar{v}A \tag{1}$$

where $\bar{v}$ is the average thermal velocity of the molecule and A is the total area of the pump where particles can enter.

For most of the calculations the source of the particles is at the entrance to the neutraliser. When considering the flow of gas from MAST into the beamline (see Section 4.6) MAST was represented by a large spherical cell at the end of the duct with it acting as the source of the torus gas. Since the particles are non-interacting, the set energy can be arbitrary. In each run 200,000 particles are tracked.

### 4.3 Interpretation of the MCNP results

Cells have been included in the model to allow the pressure in the space occupied by the cell to be calculated. These cells are along the beam path from the exit of the neutraliser to the entrance to MAST. Two cells have also been added in the region next to the walls at the Penning gauge locations. These cells allow the pressure to be calculated for comparison with the measured pressure. They had dimensions 30x30x30cm$^3$. It was checked that the results were independent of the cell size.

The MCNP tally F4 is used to calculate the gas density in a cell. This tally is known as the track length estimate of cell flux [9]. It sums the track length of each particle passing through the cell divided by the cell volume which gives the units of flux. In the MCNP output it is reported for a unit cell volume and all tallies are normalised to the number of starting particles. Its value is then the mean track length in the cell, MTL, i.e.

$$\text{F4 Tally} = \frac{1}{NV}\sum l = \frac{1}{N}\sum l = MTL \tag{2}$$

where l is the length of the path of a particle passing through the cell, N is the number of particles in the problem and the cell volume V = 1. The mean particle residence time in the cell, $\tau$, can then be calculated from

$$\tau = \frac{MTL}{\bar{v}} = \frac{V}{Q_V} \tag{3}$$

where $\bar{v}$ is the average thermal velocity, V is the cell volume and $Q_V$ is the volumetric flow rate. In terms of the net gas throughput $Q_{net}$ this becomes

$$\frac{MTL}{\bar{v}} = \frac{VnkT}{Q_{net}} \tag{4}$$

n being the density in the cell, k is Boltzmann's constant and T the temperature. The density and pressure in the cell, P = nkT can then be calculated. Finally since we will be interested in re-ionisation, the cell target density $\Pi = nL$ where L is the length of the cell along the beam path.

In equation 4 the net gas throughput is the total throughput to the ion source and neutraliser, Q, adjusted for the gas throughput associated with the neutral beam injected into the tokamak $Q_b$ i.e. $Q_{net} = Q-Q_b$. Knowing the powers of the full, half and third energy components ($P_1$, $P_{1/2}$, $P_{1/3}$ respectively) of the neutral beam this correction can be estimated. For a neutral beam at energy E (keV) the equivalent neutral beam "current", $I_{eq}$ (A) is

$$I_{ea} = 1000\left(\frac{P_1}{E} + \frac{P_2}{E/2} + \frac{P_3}{E/3}\right) \tag{5}$$

This "current" represents a particle loss rate to MAST of $I_{eq}/(2e)$ molecules/s where the factor of 2 is used to convert from atoms to molecules and e is the electronic charge. This can then be converted to a gas flow rate due to the beam, $Q_b$,

$$Q_b = 10\left(\frac{I_{eq}}{2e}\right)kT \ (mbarl/s) \tag{6}$$

For 75kV, 65A deuterium beams then $Q_b$ is approximately 5.6 mbarl/s for supercusp and 7.3 mbarl/s for chequerboard ion source configurations. For asynchronous operation where no beam is injected into MAST beams then $Q_{net} = Q$. In this case there is a source of gas due to the neutral beam in the calorimeter tank but the pumping calculations do not take this into account at present.

**4.4 Benchmarking the use of MCNP**

The use of the MCNP code has been benchmarked against a relatively simple example of a pipe with a length to radius (L/r) of 100 with r=10cm. Gas is introduced at one end of the pipe with a throughput of 10 mbar l/s and the gas is pumped at the other end. The pumping speed is given by equation 1 and for $D_2$ at 300K this is ~9870 l/s. This simple geometry has an analytical solution where the pressure along the pipe, P(l), is given by

$$P(l) = \frac{Q}{C(l)} \tag{7}$$

where Q is the gas throughput and C(l) is the conductance of the pipe of length l where l is measured from the pump. This conductance is given by [10]

$$C(l) = \frac{A\bar{v}}{4}K(l/r) \tag{8}$$

where A is the cross-section area of the pipe and K(l/r) is the Clausing factor. In addition the MCNP results have been benchmarked against another Monte-Carlo code used for vacuum calculations: MolFlow+ [11]. (The author was not aware of this code until most of the work reported here had been completed.) The results of the benchmark problem are shown in Figure 8 where the three methods of calculation all show very good agreement.

**4.5 Calculation of the pumping speed**

The speed of each TSP in both the magnet and calorimeter tanks was calculated using MCNP. The method used was to adjust the set speed for each pump by adjusting the width of the four chambers of each pump in the model. The width was equal for each chamber in each of the two tanks. The speed was adjusted until the pressures calculated in the cells using MCNP both agreed with the measured peak pressures to better than 1%. This required a simple iterative process and needed usually four or five runs to converge on the answer.

In Figure 9 and 10 the calculated pumping speed associated with the peak pressures shown in Figure 5 and 6 are shown for the 1.3s and 6.5s duration gas pulses. For the 1.3s gas pulses (Figure 9) the pumping speeds in both tanks fall with each gas pulse until the evaporations begin. Then they increase when all chambers are evaporated, remain constant while one chamber is evaporated and increase again when two chambers are evaporated. However there is, in this case a large difference in the speeds of the two pumps when the evaporations are under way. For the 6.5s duration gas pulses, although the speed starts out differently they are relatively close to one another through the evaporation regimes but both pumping speeds are lower than for the 1.5s gas pulses. The MCNP particle accounting shows that the fractions of gas pumped in the magnet and calorimeter tanks for the 1.3s gas pulses is in the ratio ~0.87/0.13 and for the 6.5s gas pulses is ~0.84/0.16. These values are effectively constant for all evaporation regimes. This is due to the intertank duct between the magnet and calorimeter tanks as seen in Figure 1. This has an estimated conductance of ~ 14,600 l/s for $D_2$ at 300K. Thus it will dominate the effective pumping speed for gas to enter the calorimeter tank compared to being pumped in the magnet tank.

Without knowing the details of the surface conditions curves such as these can be difficult to interpret. The highest pumping speed measured was ~160,000 l/s. Feist et al report a maximum specific pumping speed for this pump of 3-4.5 l/s/cm$^2$ for hydrogen. From a simple estimate of the surface area, using a factor of 3 to account for the corrugation of the surface [8], this speed corresponds to a specific pumping speed of ~3.4 l/s/cm$^2$. The actual pumping surface area may be higher since the evaporation deposits titanium on areas outside the pump as the chambers are not closed.

The pump in the magnet tank shows a higher pumping speed than that in the calorimeter tank. The lower pumping speed at the higher pressures associated with the longer gas pulse lengths for both is to be expected [8,12] but the decrease is larger for the magnet tank pump compared to the calorimeter tank pump. These features may be associated with a number of causes. At the lower pressure in the calorimeter tank, effects of any leaks which could reduce the pumping speed would be more significant. At the lower pressure in the calorimeter tank the pump might be operating closer to the titanium limited and surface conductance limited regime transition [12]. The change in pumping speed with pressure would be less than in the titanium limited regime at higher pressures. It was found however that the pump in the magnet tank was operating at ~88-90A for evaporation whereas the calorimeter tank pump was operating at ~80A. This means that the amount of titanium evaporated onto the calorimeter tank pump would have been less than for the magnet tank. The filaments in the magnet tank pumps had an integrated usage which was higher by ~30% than the calorimeter filaments and this might affect the amount of titanium evaporated also.

### 4.6 Application to re-ionisation calculations

The methods discussed so far can be used in the case of injection of a neutral beam into MAST. Figure 11 shows the pressures (on a linear scale) measured in the magnet and calorimeter tanks of the South beamline during MAST pulse 29061. This pulse had an extracted ion beam of 55.8A at 73kV for a duration of 0.3s, resulting in an injected neutral beam power of 2.05MW. As illustrated in Figure 11 the pressure in the beamline falls on extraction of the beam, reaches a minimum value and then recovers somewhat at the end of the beam pulse. The pressure is a balance of a number of gas sources: the gas flow from the ion source and neutraliser, the gas flow represented by the neutral beam which is injected into MAST, pumping and subsequent re-release as gas of the residual positive ions in the ion dumps and any gas flow from the torus into the beamline. The gas pressure in the beamline has been calculated using the MCNP model taking into account the gas flow of the neutral beam injected into MAST and gas from the torus. Any residual ion effects have not been modelled since this represents a gas source in the magnet tank where most of the gas is pumped. The pressures at the end of the pulse have been used in the calculation shown here as these values are close to being an average for the pulse.

In its present configuration the maximum gas density on the outside of MAST is of the order of $2 \times 10^{18}$ m$^{-3}$ ($8.2 \times 10^{-3}$ mbar). In the MCNP model, MAST is represented as a sphere of volume $V_{MAST}$ at the end of the duct acting as a gas source. The particle flow rate from MAST, $Q_{MAST}$, can be calculated from the mean track length $MTL_{MAST}$ for the MAST cell at the required density, $n_{MAST}$, and average thermal velocity, $\bar{v}$, by

$$Q_{MAST} = \frac{n_{MAST} \bar{v} V_{MAST}}{MTL_{MAST}} \ (particles/s) \qquad (9)$$

The density in any cell can then be calculated from equation (4) with $Q_{net}/kT$ replaced by $Q_{MAST}$.

In order to carry out the calculations MCNP was run and analysed for the case of source and neutraliser gas only taking into account the gas flow represented by the neutral beam injected into MAST. The total gas flow from the source and neutraliser was 30 mbar l/s and the neutral beam was an equivalent to a gas throughput of 4.4 mbar l/s. It was also run for the case of gas from MAST. The pressures in the cells from both runs were added together and the pumping speeds then adjusted to match the total pressures in the cells representing the Penning gauges. The gas profiles calculated for MAST pulse 20961 are shown in Figure 12 plotted as a function of distance from the grounded grid in the accelerator. At this level of gas in the torus, the contribution from the torus to the pressure in the duct region is significant.

The pumping speeds in the case of no torus gas were 151,600 l/s and 60,800 l/s in the magnet and calorimeter tanks respectively. For the case with additional torus gas these speeds were 152,100 l/s and 69,500 l/s respectively. There is only a small change in the magnet tank speed to account for the additional gas from the torus but a much larger change in the calorimeter tank pump speed since most of the gas is pumped here and the speed increases to maintain the same measured pressure. Of the source and neutraliser gas, after accounting for the neutral beam, ~88% is pumped in the magnet tank, 10% in the calorimeter tank and 2% enters MAST. Approximately 80% of the torus gas is pumped in the calorimeter tank and the remainder in the magnet tank. The re-ionisation can then be simply calculated using the gas profiles and the appropriate cross-sections [13]. Account is taken of the full, half and third energy fractions of the neutral beam power (85.2%, 11.9%, 2.9% respectively) to give the total re-ionisation as a fraction of the neutral beam power. The results are summarised in shown in Figure 13 which gives the re-ionisation from the entrance of the intertank duct downstream and also in the duct region alone. As expected the re-ionisation within the beamline is lower for the case of no torus gas. The overall levels of re-ionisation with torus gas are relatively high but this level of torus gas represents a maximum value. This should be reduced further with the use of the Super-X divertor being installed on MAST [14].

## 5. Conclusions

In this paper, titanium evaporation regimes have been found which lead to a constant pumping speed in the MAST NBI titanium sublimation pumps for short and long gas pulses. Operation with longer gas pulses requires more titanium evaporation to maintain a constant pumping speed. Such operation will reduce the period between changes of the titanium bearing filaments. This will have an impact on the operational programme since the injectors must be vented to change the filaments. A constant pumping speed is not enough in itself, the speed itself is important since this will determine re-ionisation levels in the beamline and if these are not controlled it can lead to loss of beam transmission and possibly damage to beamline components.

It has been shown how the MNCP code can be used to calculate the pumping speeds and associated re-ionisation levels in the beamline. Provided enough titanium evaporation is carried out the pumping speed can reach acceptable levels resulting in a few percent re-ionisation which can then be maintained by the appropriate evaporation regimes depending on the pulse length. These methods are a valuable tool in understanding the beamline performance. The calculations can be improved by using the positive ions which strike the residual ion dump along with those neutral particles which strike the calorimeter as part of the gas source. This work can be advanced further by using the calculated gas profiles and to track the trajectories of the re-ionised particles created along the beamline to calculate power densities on the various components.

## Acknowledgements

The author would like to thank M Kovari for his assistance in finding my way through the use and interpretation of the MCNP code and the MAST NBI team for helping with the experiments. This work was funded by the RCUK Energy Programme [under grant EP/I501045] and the European Communities under the contract of Association between EURATOM and CCFE. To obtain further information on the data and models underlying this paper please contact PublicationsManager@ccfe.ac.uk. The views and opinions expressed herein do not necessarily reflect those of the European Commission.

# Figure Captions

Figure 1 (Colour online) Schematic of the MAST Neutral Beam Injectors

Figure 2 (Colour online) Sketch of the Titanium Sublimation Pumps in the MAST NBIs (dimensions are in cm)

Figure 3 (Colour online) Pressure traces for a gas only pulse at 30 mbar l/s

Figure 4 Peak pressures in (a) the magnet tank and (b) the calorimeter tank for a series of 3.5s duration 31 mbar l/s gas pulses for initial titanium evaporation periods of 2, 3 and 5 minutes

Figure 5 Peak pressures in the magnet (closed symbols) and calorimeter (open symbols) tanks for differing evaporation regimes between gas pulses of 1.3s duration at 30 mbar l/s throughput

Figure 6 Peak pressures in the magnet (closed symbols) and calorimeter (open symbols) tanks for differing evaporation regimes between gas pulses of 6.5s duration at 30 mbar l/s throughput

Figure 7 (Colour online) The MCNP model of the MAST NBI system

Figure 8 The pressure along a pipe of L/r=1000 as calculated using MCNP, analytically and with MolFlow+

Figure 9 Calculated pumping speed in the magnet (closed symbols) and calorimeter (open symbols) tanks for differing evaporation regimes between gas pulses of 1.3s duration at 30 mbar l/s throughput. (See Figure 5 for the pressure data.)

Figure 10 Calculated pumping speed in the magnet (closed symbols) and calorimeter (open symbols) tanks for differing evaporation regimes between gas pulses of 6.5s duration at 30 mbar l/s throughput. (See Figure 6 for the pressure data.)

Figure 11 Pressures in the magnet and calorimeters tanks during MAST pulse 20961 (73keV, 55.8A, 2.05MW neutral beam injected into MAST)

Figure 12 Calculated pressure profiles in the injector beamline for MAST pulse 29061 where there is a source and neutraliser gas throughput of 30 mbar l/s, a torus gas density of $2 \times 10^{18} m^{-3}$ and the total.

Figure13 Re-ionisation levels as a percentage of the total neutral beam power for MAST pulse 29061 The values are for downstream from the start of the intertank duct and in the duct region for the case of no torus gas and with torus gas.

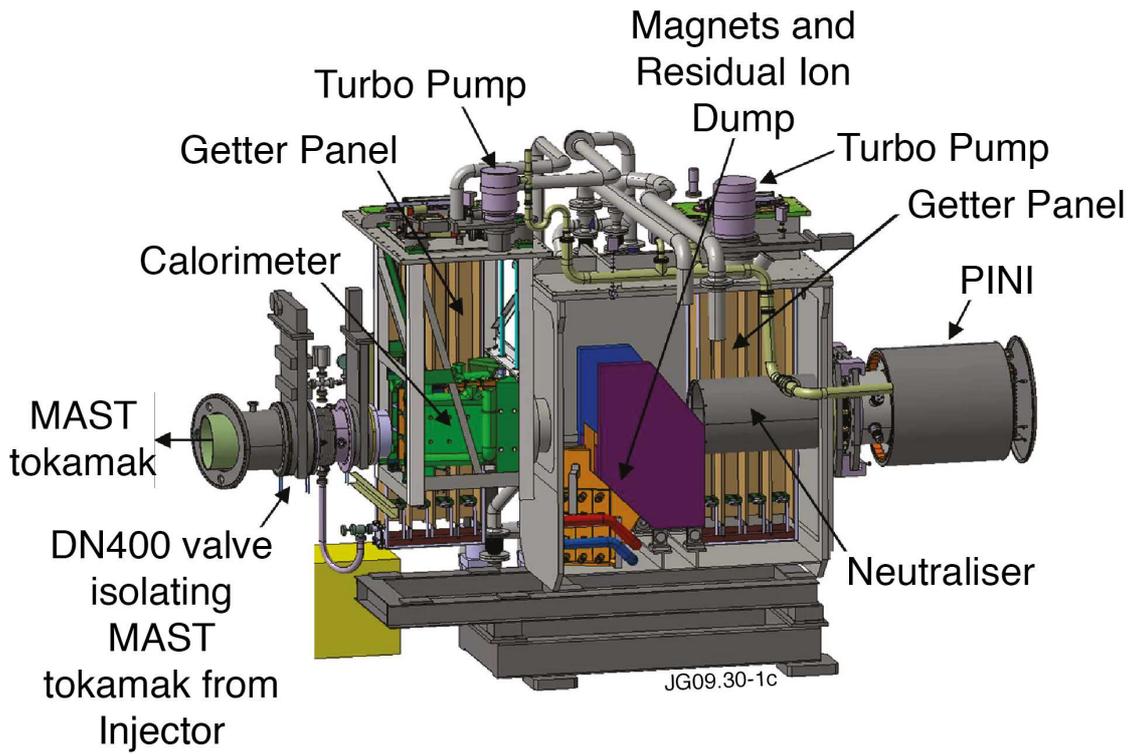

**Figure 1 (Colour online) Schematic of the MAST Neutral Beam Injectors**

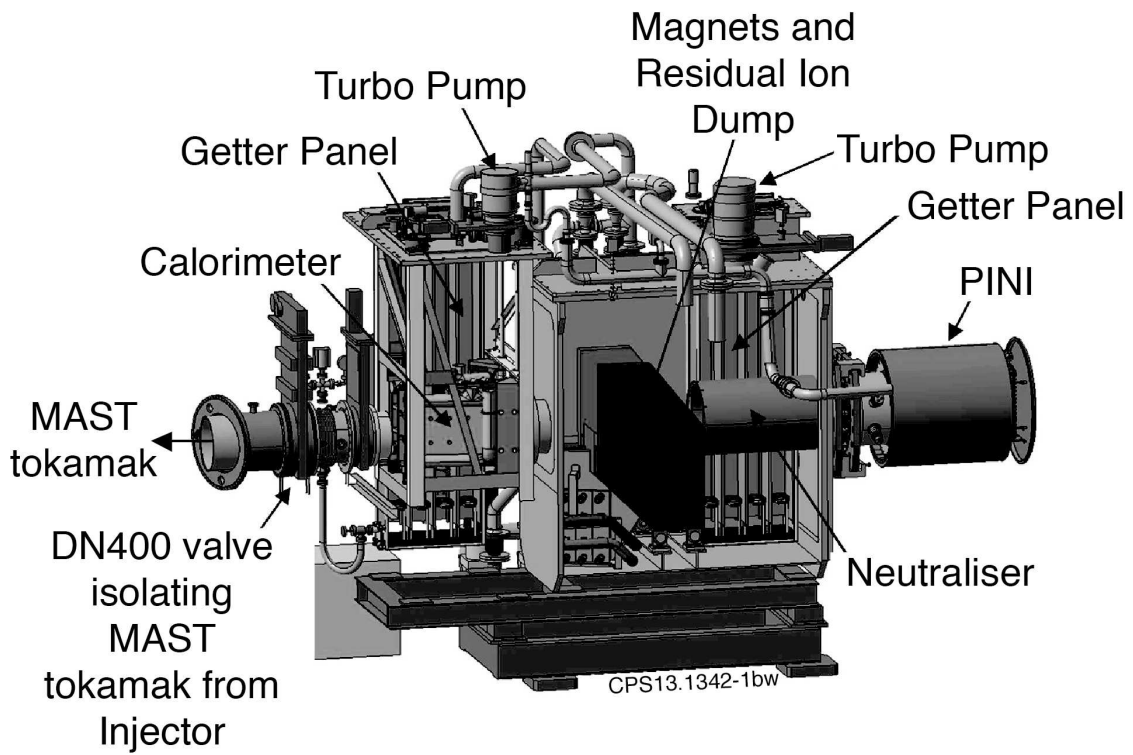

**Figure 1 Schematic of the MAST Neutral Beam Injectors**

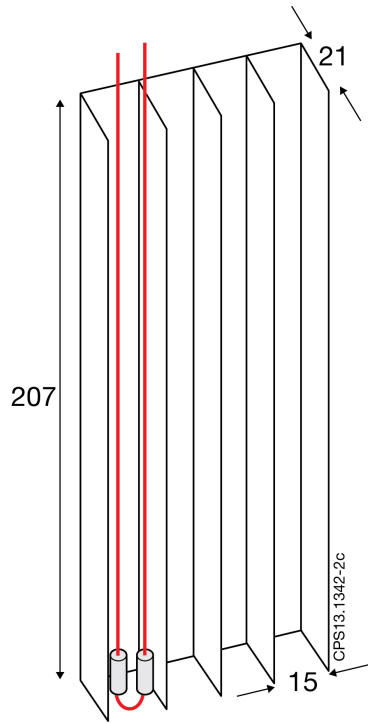

**Figure 2 (Colour online) Sketch of the Titanium Sublimation Pumps in the MAST NBIs (dimensions are in cm)**

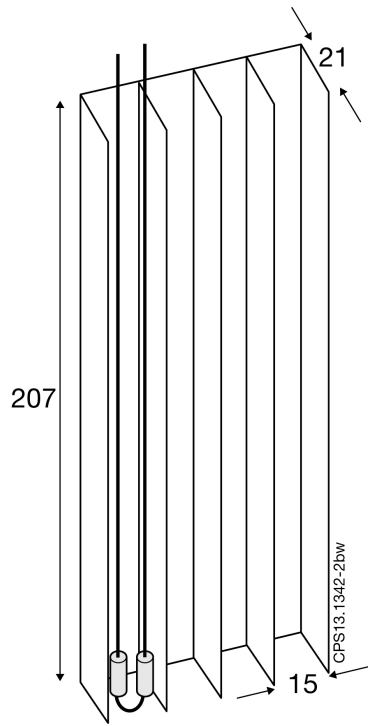

**Figure 2 Sketch of the Titanium Sublimation Pumps in the MAST NBIs (dimensions are in cm)**

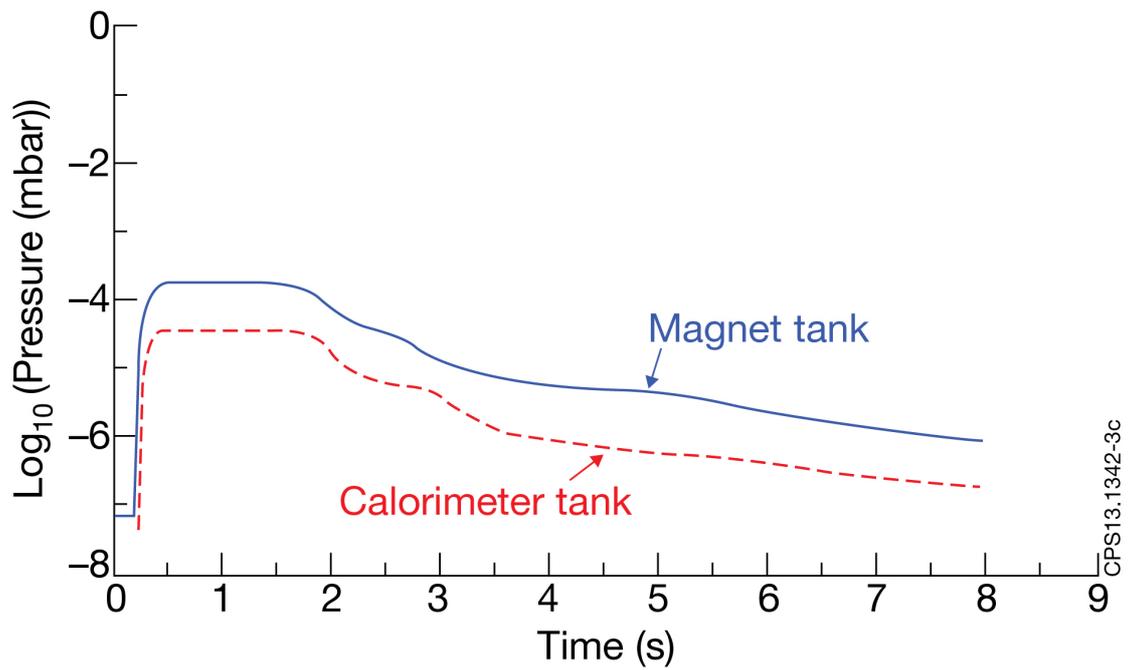

Figure 3 (Colour online) Pressure traces for a gas only pulse at 30 mbar l/s

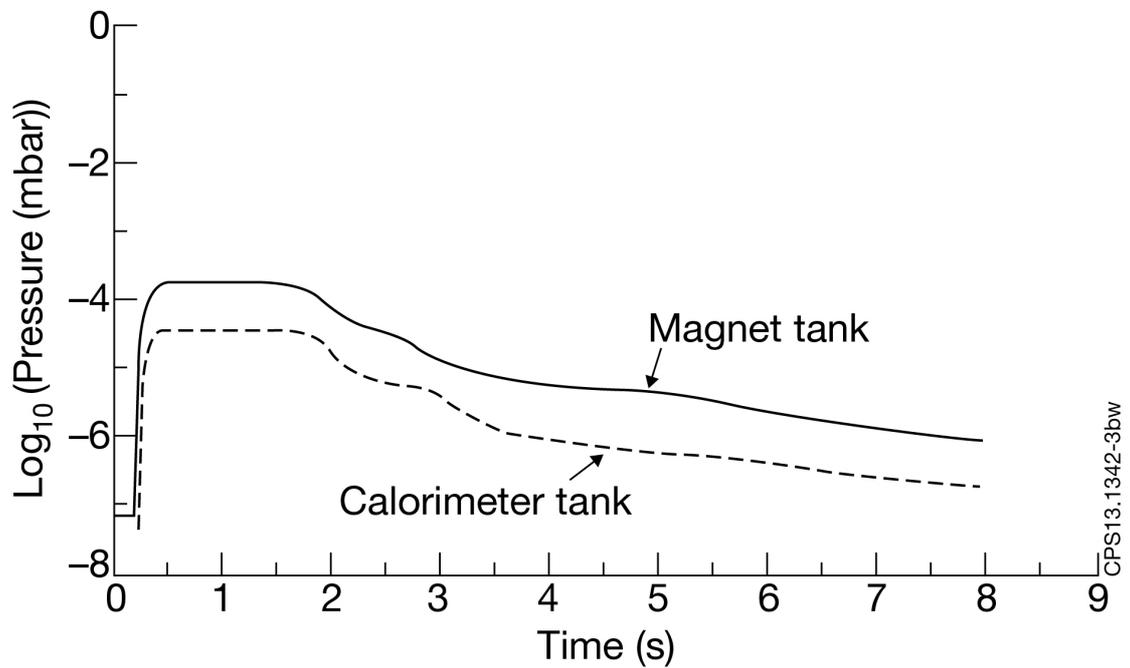

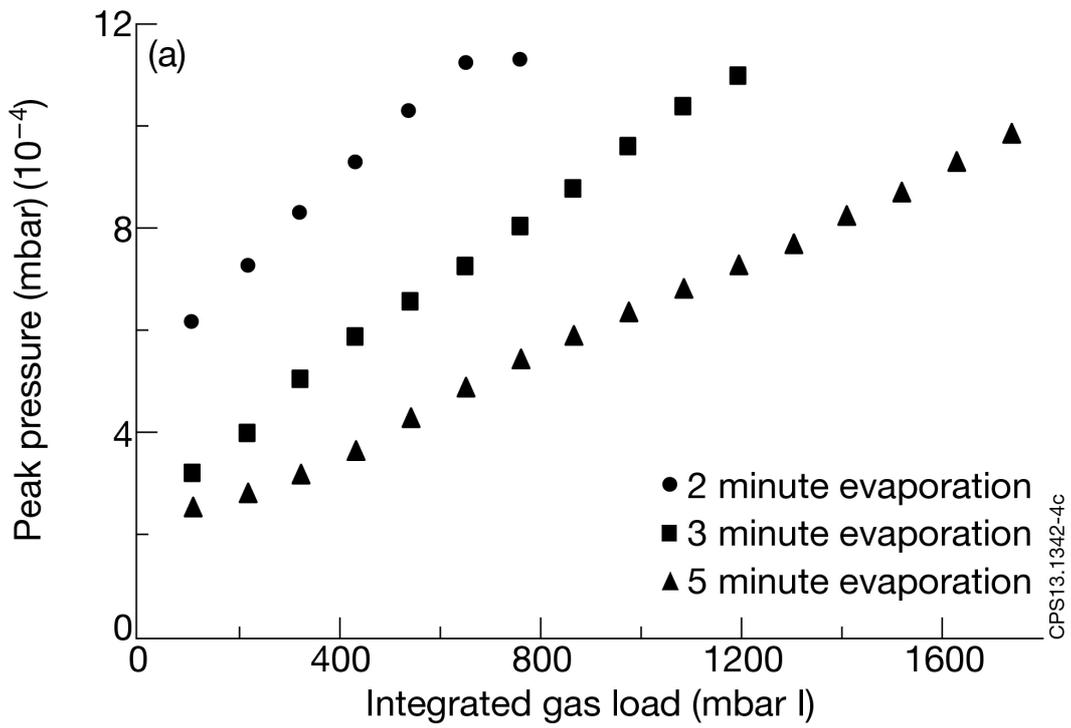

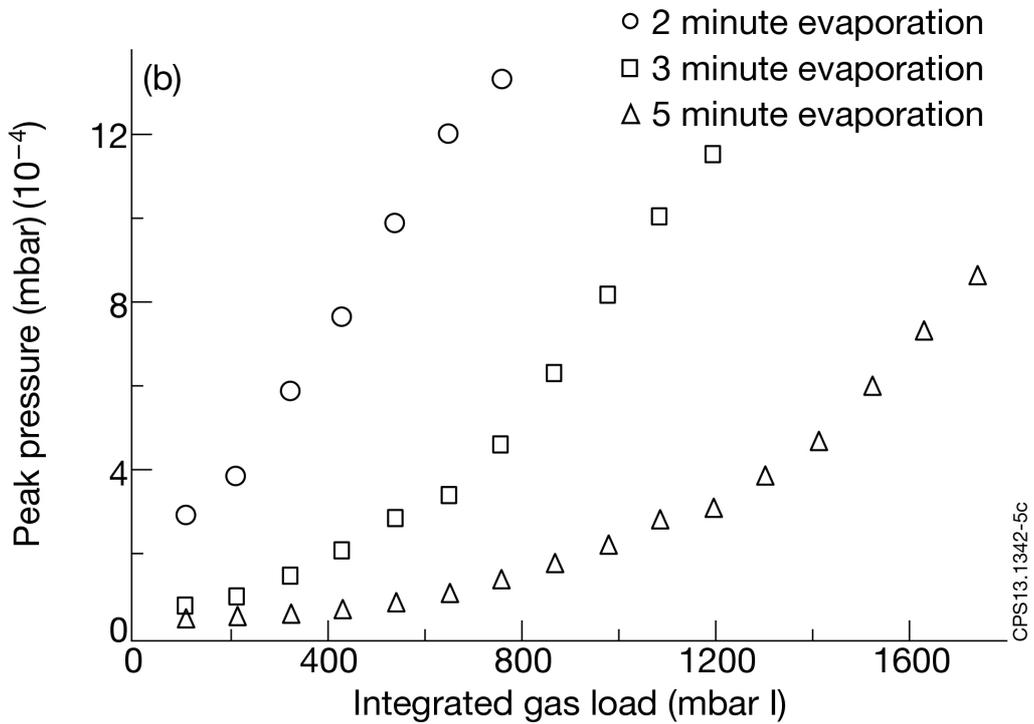

**Figure 4 Peak pressures in (a) the magnet tank and (b) the calorimeter tank for a series of 3.5s duration 31 mbar l/s gas pulses for initial titanium evaporation periods of 2, 3 and 5 minutes**

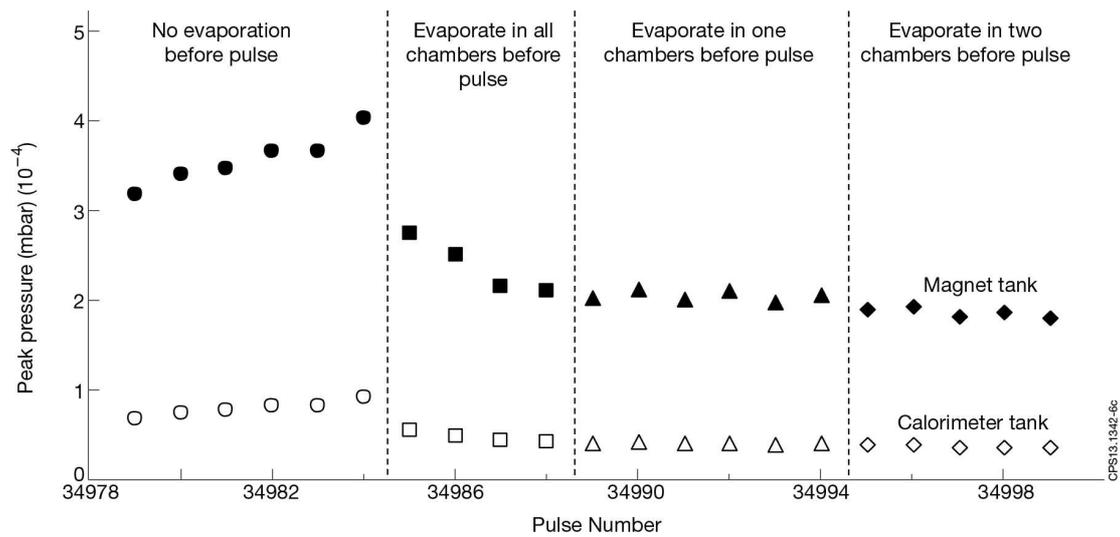

Figure 5 Peak pressures in the magnet (closed symbols) and calorimeter (open symbols) tanks for differing evaporation regimes between gas pulses of 1.3s duration at 30 mbar l/s throughput

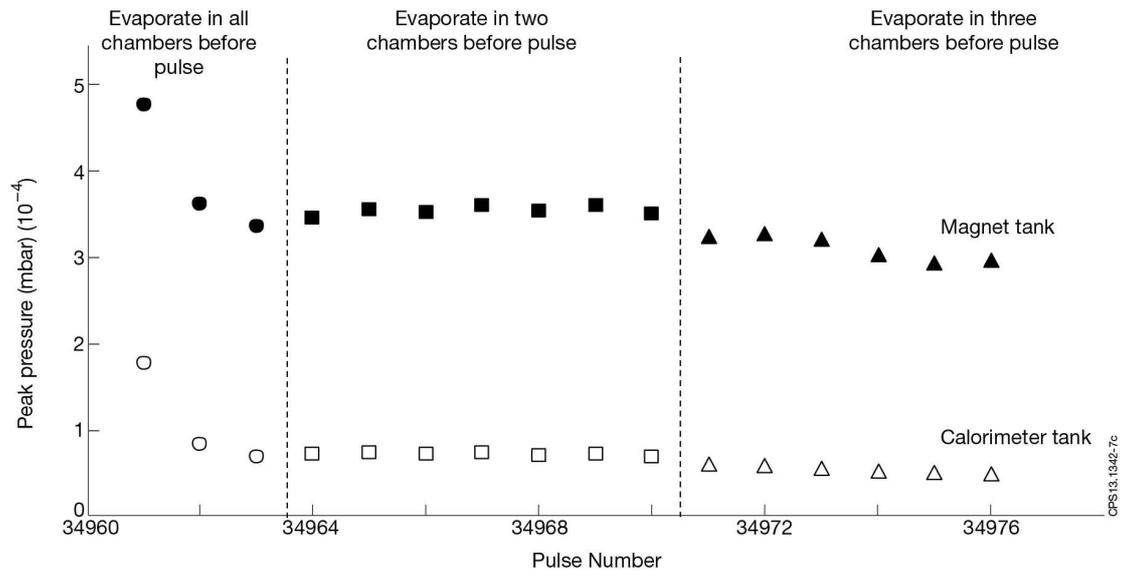

Figure 6 Peak pressures in the magnet (closed symbols) and calorimeter (open symbols) tanks for differing evaporation regimes between gas pulses of 6.5s duration at 30 mbar l/s throughput

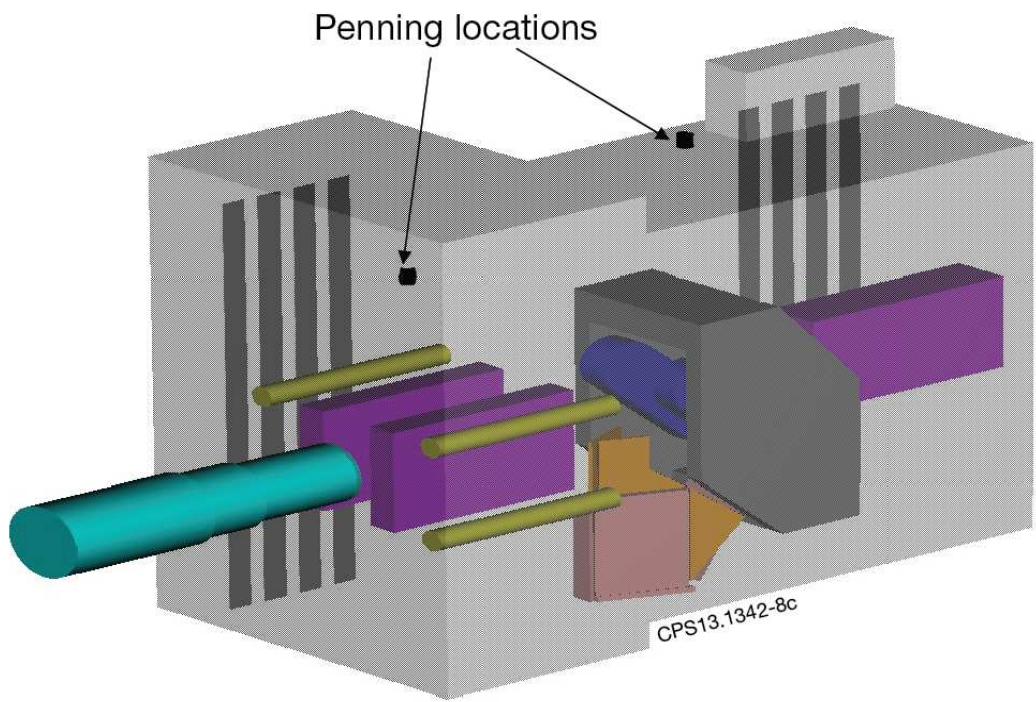

**Figure 7 (Colour online) The MCNP model of the MAST NBI system**

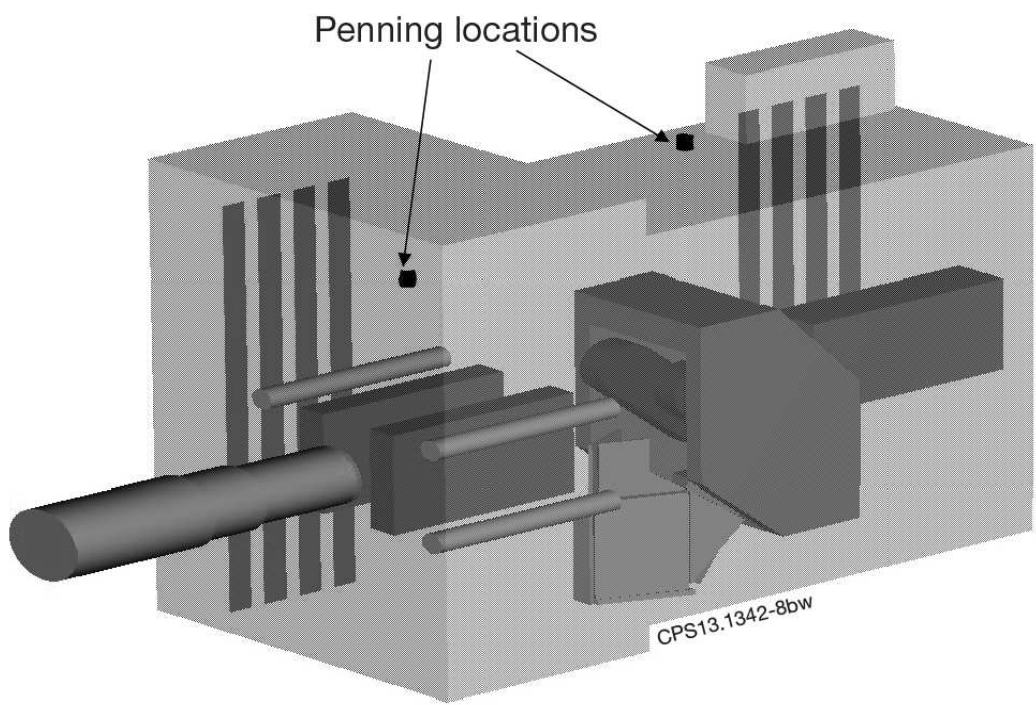

**Figure 7 The MCNP model of the MAST NBI system**

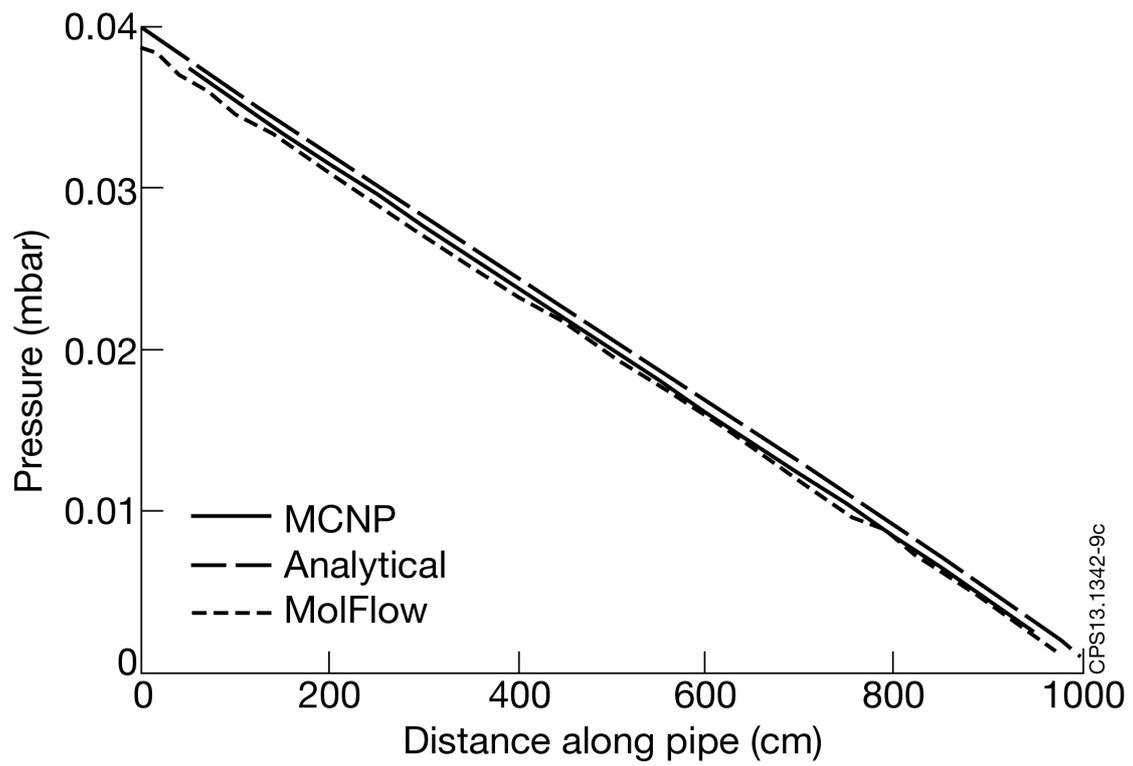

**Figure 8 The pressure along a pipe of L/r=1000 as calculated using MCNP, analytically and with MolFlow+**

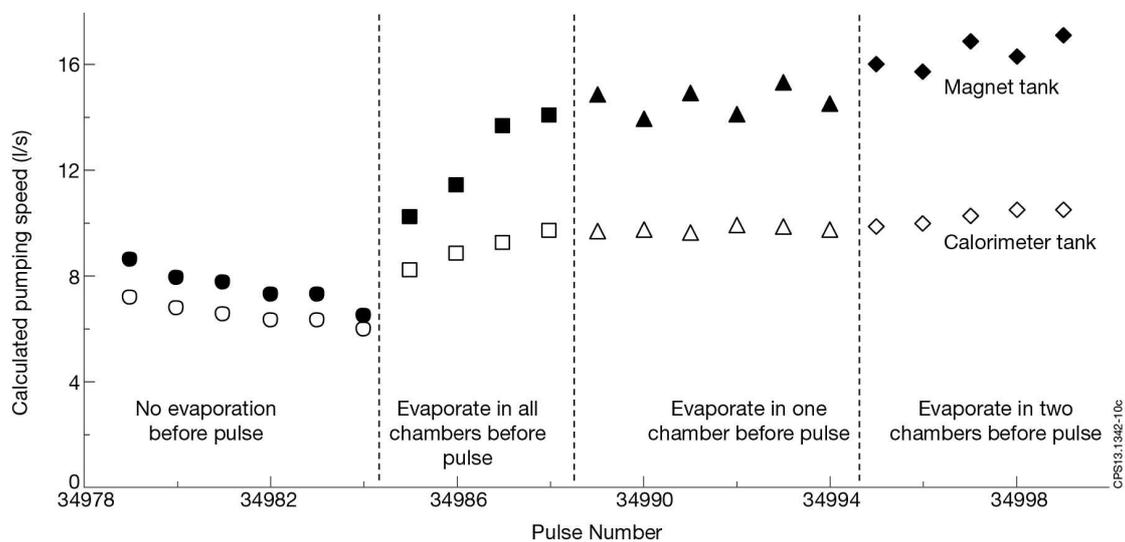

Figure 9 Calculated pumping speed in the magnet (closed symbols) and calorimeter (open symbols) tanks for differing evaporation regimes between gas pulses of 1.3s duration at 30 mbar l/s throughput. (See Figure 5 for pressure data.)

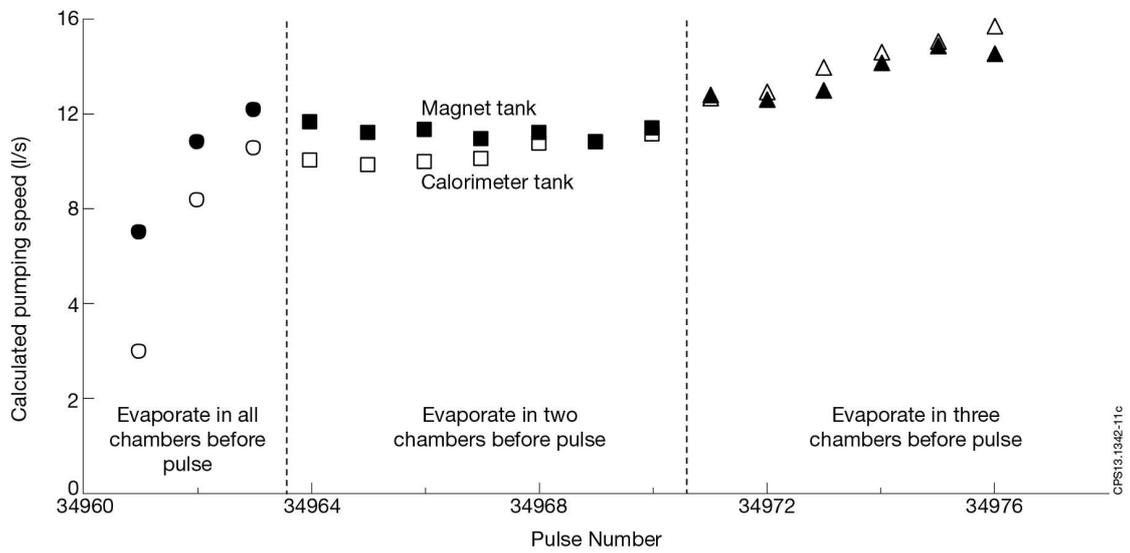

Figure 10 Calculated pumping speed in the magnet (closed symbols) and calorimeter (open symbols) tanks for differing evaporation regimes between gas pulses of 6.5s duration at 30 mbar l/s throughput. (See Figure 6 for pressure data.)

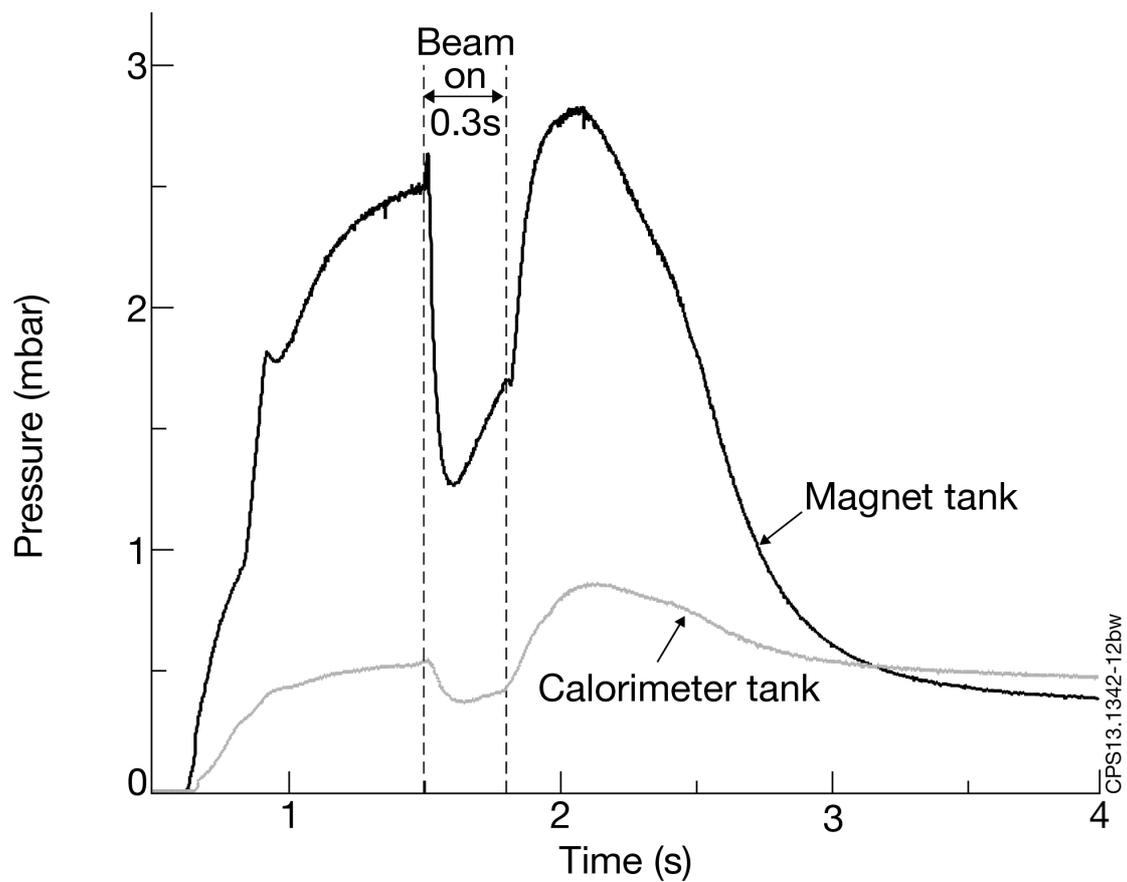

Figure 11 Pressures in the magnet and calorimeters tanks during MAST pulse 20961 (73keV, 55.8A, 2.05MW neutral beam injected into MAST)

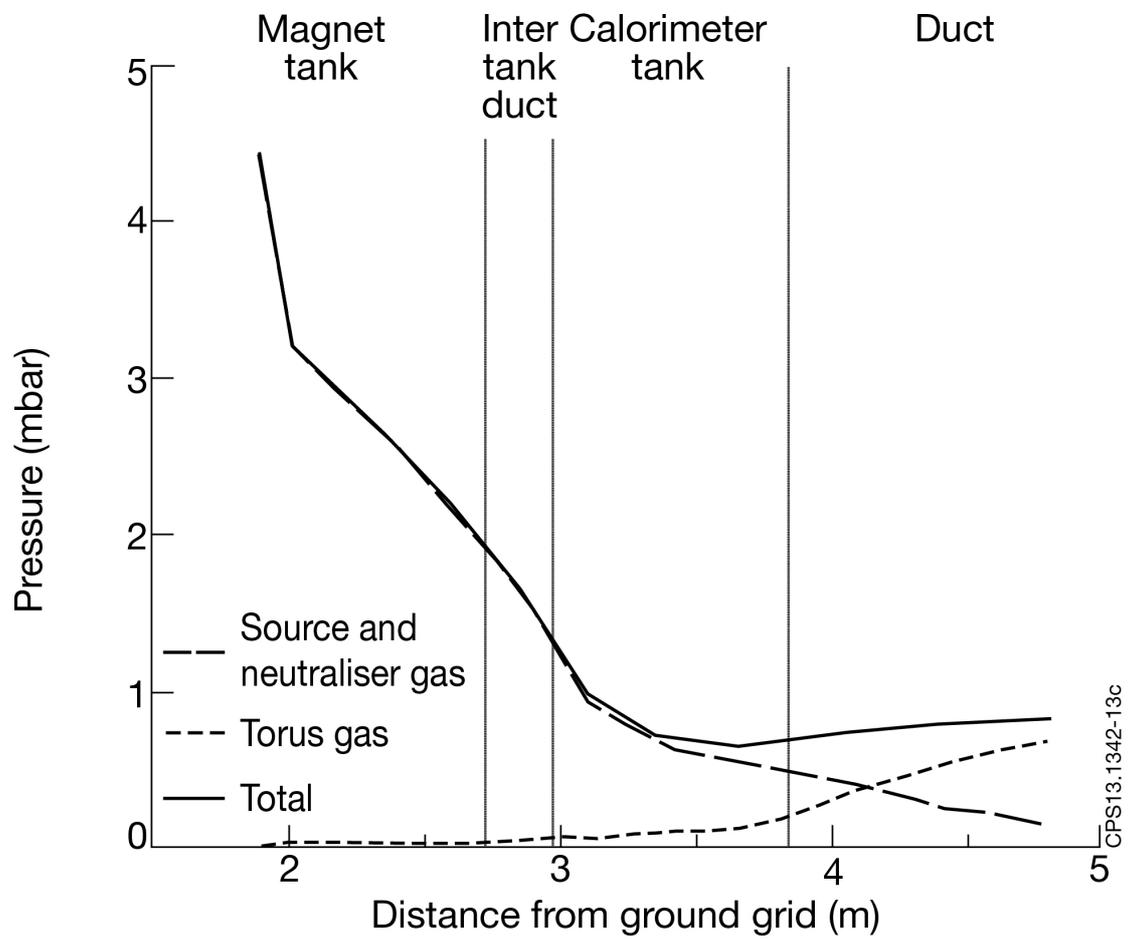

Figure 12 Calculated pressure profiles in the injector beamline for MAST pulse 29061 where there is a source and neutraliser gas throughput of 30 mbar l/s, a torus gas density of $2 \times 10^{18} m^{-3}$ and the total.

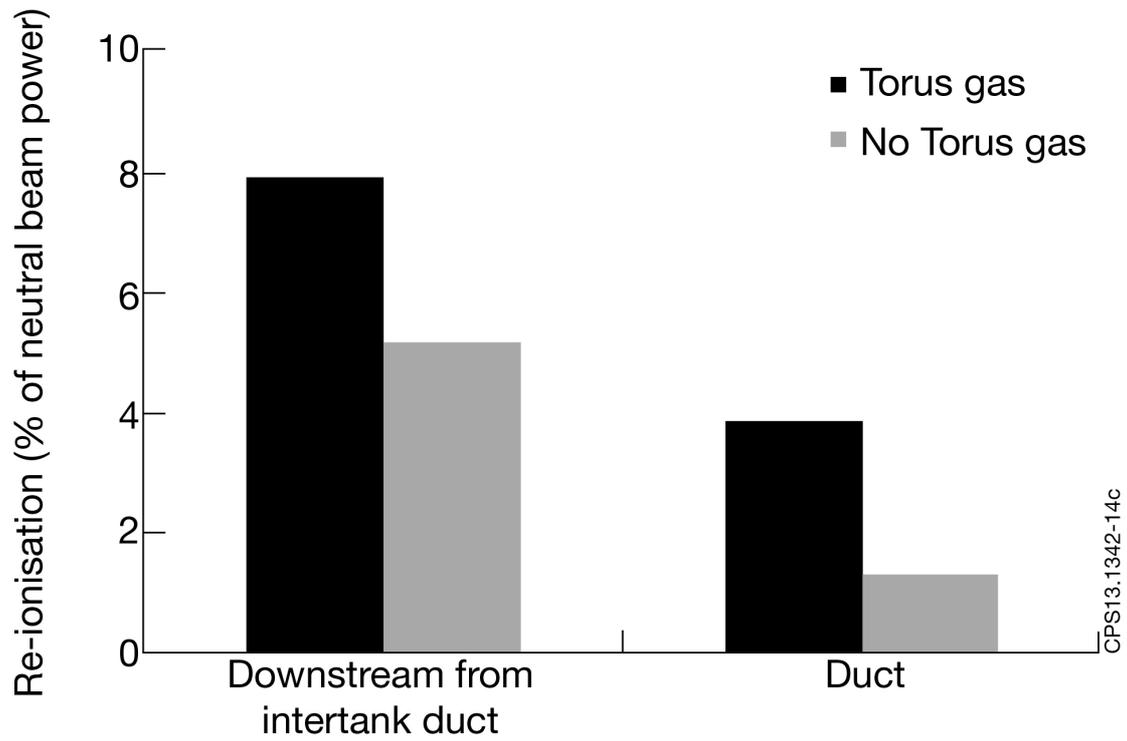

Figure13 Re-ionisation levels as a percentage of the total neutral beam power for MAST pulse 29061. Values are for downstream from the start of the intertank duct and in the duct region for the case of no torus gas and with torus gas.